\documentclass[fleqn,usenatbib]{mnras}

\usepackage[T1]{fontenc}
\usepackage{ae,aecompl}

\usepackage{graphicx}	
\usepackage{amsmath}	
\usepackage{amssymb}	

\usepackage{hyperref}
\usepackage{cleveref}
\usepackage{booktabs}
\usepackage{multirow}
\usepackage[dvipsnames]{xcolor}
\usepackage[normalem]{ulem}

\definecolor{darkgreen}{rgb}{0.1, 1.0, 0.5}

\newcommand*{\ksM}{\text{km/s Mpc$^{-1} $}}

\newcommand{\Al}{A_{\rm lens}}
\newcommand{\Alp}{A_{\rm lp}}
\newcommand{\nlp}{n_{\rm lp}}
\newcommand{\tr}{\tau}
\newcommand{\LCDM}{$\Lambda$CDM }
\newcommand{\rd}{r_{\rm d}}
\newcommand{\fEN}{f_{\rm EDE/ NEDE}}
\newcommand{\fE}{f_{\rm EDE}}
\newcommand{\fN}{f_{\rm NEDE}}

\defcitealias{Riess:2020fzl}{R20}
\defcitealias{Chudaykin20}{C20}

\title[]{Scrutinizing Early Dark Energy models through CMB lensing}

\author[B. S. Haridasu et al.]{
Balakrishna S. Haridasu,$^{1,2,3,4}$\thanks{sandeep.haridasu@sissa.it}
Hasti Khoraminezhad,$^{1,2,4,5}$\thanks{viel@sissa.it}
Matteo Viel$^{1,2,3,4}$\thanks{hk3b3@mst.edu}\\
$^{1}$SISSA-International School for Advanced Studies, Via Bonomea 265, 34136 Trieste, Italy\\
$^{2}$INFN, Sezione di Trieste, Via Valerio 2, I-34127 Trieste, Italy\\
$^{3}$INAF - Osservatorio Astronomico di Trieste, Via G. B. Tiepolo 11, I-34143 Trieste, Italy\\
$^{4}$IFPU, Institute for Fundamental Physics of the Universe, via Beirut 2, 34151 Trieste, Italy\\
$^{5}$Institute for Multi-messenger Astrophysics and Cosmology, Department of Physics, \\Missouri University of Science and Technology, 1315 N Pine St, Rolla, MO 65409, USA
}

\date{Accepted XXX. Received YYY; in original form ZZZ}

\pubyear{2021}

\begin{document}
\label{firstpage}
\pagerange{\pageref{firstpage}--\pageref{lastpage}}
\maketitle

\begin{abstract}
We investigate early dark energy models in the context of the lensing anomaly by considering two different Cosmic Microwave Background (CMB) datasets: a complete Planck, and a second one primarily based on SPTPol and Planck temperature ($l<1000$). We contrast the effects of allowing the phenomenological lensing amplitude ($\Al$) to be different from unity. We find that the fraction of early dark energy, while not immediately affected by the lensing anomaly, can induce mild deviations, through correlations with the parameters  $H_0$ and $S_8$. {We extend the analysis also by marginalizing the Newtonian lensing potential, finding a $\gtrsim 1\sigma$ deviation, when allowing for an amplitude rescaling and scale-dependence. Modeling the rescaling of the theory lensing potential and the acoustic smoothing of the CMB spectra, we find that only to a moderate level the anomaly can be addressed by modifying the lensing signal itself and that an additional $\Al \sim 1.1$ at $\sim 2\sigma$ significance should be addressed by pre-recombination physics.  Finally, we also comment on the lensing anomaly in a non-flat ($\Omega_{\rm k} \neq 0$) scenario, finding that the late-time flatness of the universe is robust and not  correlated with the additional smoothing in the CMB spectra.}

\end{abstract}




\section{Introduction}

In defiance of the achievements of the concordance \LCDM model, several tensions have gained significance in recent years. Among them, the so-called $H_0$-tension \citep{Verde:2019ivm, Riess:2020fzl, Soltis:2020gpl, DiValentino:2020zio}, which is a disagreement between the local measurement of the current expansion rate using Type-Ia supernovae \citep{Riess19,Riess:2020fzl, Riess:2021jrx} and the indirect model-dependent estimate inferred from the Cosmic Microwave Background (CMB), primarily by Planck \citep{Aghanim:2018eyx}, and to a mild degree in other observations \citep{ Hinshaw_2013, ACT:2020gnv, SPT-3G:2021eoc}, has now risen to a $\sim 4.2\sigma$ significance. This tension, once other late-time estimates \citep{Wong19,Birrer20} are considered, has also been reported at $\sim 6 \sigma$. The second tension, also gaining notable significance is for the determination of $S_8 \equiv \sigma_8 (\Omega_m/0.3)^{0.5}$, now reaching $\sim 2-3\sigma$ level \citep{Joudaki:2019pmv, Heymans20,DES:2021wwk, DiValentino:2020vvd}. These tensions should however be treated in conjunction, without worsening one when alleviating the other (see for example \citep{Ivanov:2020ril, Smith:2020rxx, Jedamzik:2020zmd}).


Assuming these tensions are not due to the unknown systematics within observations, numerous models have been proposed and explored to resolve the $H_0$-tension: i) modifying the dark matter phenomenology \citep{Hryczuk20, Vattis:2019efj, Anchordoqui:2020djl, Archidiacono19, Blinov:2020uvz,  Haridasu:2020xaa}, ii) various dark energy scenarios \citep{Karwal16, Poulin18, Poulin18a, Hill:2020osr, Ivanov:2020ril, Smith:2020rxx, Yang18a, Pan19, DiValentino:2017iww, Khosravi17, xiaviel09, Khoraminezhad:2020cer, Ye:2020btb, Ye20b, Lin19, Lin20,  Mortsell18, Das20, Sakstein19, Niedermann:2019olb, Niedermann:2020qbw, Lin20, Fondi:2022tfp, Cruz:2022oqk}. Amongst these different proposals, modifications of the early universe physics could provide more plausible scenarios once a statistical comparison with other models is performed \citet{Schoneberg:2021qvd} (see also \citep{Moss:2021obd} for a reconstruction-based analysis).

To this extent, it is of primary importance to assess the viability of these models in several scenarios and using different combinations of datasets, in particular Planck \citep{Planck18_parameters} and SPTPol \citep{SPT:2017jdf}. Very recently a similar comparison was made between the Planck and ACTPol \citep{ACT:2020gnv} CMB datasets \citep{Hill:2021yec, Poulin:2021bjr}, both showing that the joint analysis with Planck strongly reduces the evidence for EDE. In essence, the Planck CMB data do not show  a clear evidence for EDE models, as was also noted in \citep{Haridasu:2020pms}. In this work we test the early dark energy (EDE) and the New EDE (NEDE) models using two different CMB data combinations, trying to assess some of the correlations with the standard cosmological parameters and possible subtle effects one should look for. The goal of these proposals is to allow for larger values of $H_0$ than the one derived in the \LCDM model. This would be achieved by demanding that the angular acoustic scale (the ratio of the sound horizon at last scattering to the comoving angular diameter distance) remain unchanged by the new physics introduced to alleviate the $H_0$ tension.  In this context, we assess the limits on the fraction of early dark energy while accounting for the CMB lensing anomalies, which vary between CMB datasets. Essentially, we take into account the fact that \textit{Planck} CMB dataset predicts a phenomenological lensing amplitude $\Al > 1$ \citep{Aghanim:2018eyx}.  More recently, \cite{Fondi:2022tfp}, have investigated the same in the context of \textit{Planck} CMB dataset and showed that there is no evidence for an early dark energy model when taking into account the lensing anomaly. Here we extend the same, to an additional NEDE model and also consider an additional CMB dataset mainly based on SPTPol. 

Moreover, we also perform the early universe analysis (EUA) as elaborated in \citep{Haridasu:2020pms}, wherein we adopt the methodology presented in \citep{Verde17}, which was implemented to obtain constraints only utilizing the information from the early universe within the CMB dataset. Earlier we used this approach in \citep{Haridasu:2020pms}, to show that when EDE is assessed within the EUA, there is a mild reduction in limits of the allowed dark energy fraction at early times. We now extend this to both models using the two CMB datasets. The primary necessity in such an analysis is to marginalize the lensing effects which are late-time effects while obtaining constraints only derived from the early universe. Going through the procedure we also report some tentative yet potentially important deviations for the CMB lensing analysis. Finally, we use this analysis to comment on the correlations between different lensing parameters. In \citep{Haridasu:2020pms}, we also noted that an early-time modification alone might not be sufficient in addressing the $H_0$-tension, which was also suggested in several other works \citep{Jedamzik:2020zmd,Lin:2021sfs, Vagnozzi:2021tjv, Bernal:2021yli} through very detailed analyses, in which consequences in terms of several dynamical and geometrical observables of the large scale structure are also explored in great detail

The current paper is organized as follows: In \Cref{sec:Theory} we briefly describe the theoretical models and the datasets. In \Cref{sec:Lensing} we describe the modifications to the lensing analysis performed. The results and discussions are presented in \Cref{sec:Results} and finally we conclude in \Cref{sec:Conclusions}. We show the contour plots and some extended discussions in \Cref{sec:CommentC20} through \Cref{sec:Tri_figures}.

\section{Theory}
\label{sec:Theory}
In this section we briefly introduce the models, the datasets used, and finally, the likelihood analysis performed. 
\subsection{Early dark energy models}

As already mentioned, we compare and contrast two competing early dark energy models, essentially introduced to alleviate the $H_0$-tension. These models allow a higher value of $H_0$, which simultaneously decreases the sound horizon at drag epoch ($\rd$), keeping the combined value $H_0\times \rd$ constant. We keep the introduction to the theoretical modeling to a minimum level as they have been established extensively in the literature. {The Early Dark Energy model\footnote{The EDE implementation in \texttt{CLASS}, \texttt{CLASS\_EDE} \citet{Hill:2020osr} is available at \href{https://github.com/mwt5345/class_ede}{https://github.com/mwt5345/class\_ede}.} implements a single axion-like light scalar field allowing  the effective cosmological constant to dynamically decay. The potential of the scalar field $\phi$ is given as:} 

\begin{equation}
    V(\phi) = m^2f^2(1-cos(\phi/f))^n \,. 
\end{equation}
{The physics of the scalar field can be described through effective parameters $z_{c}$, $f_{\rm EDE} = \rho_{\rm EDE}(z_{\rm c})/3m_{\rm pl}^2H(z_{\rm c})^2 $ and $\theta_{\rm i} = \phi_{\rm i}/f$. Here $z_{\rm c}$ denotes the redshift at which the EDE contributes the most to the total energy density (please refer to \citet{Poulin18, Poulin:2018dzj, Hill:2020osr, Ivanov:2020ril} for an extensive description of the theory and the modeling). Therefore, the effective dynamics of the scalar field will be described by three additional parameters $\Theta_{\rm EDE} \equiv \{f_{\rm EDE}, \log_{10}(z_{\rm c}), \theta_{\rm i}\}$.} In the current analysis we fix the parameters $\{ \log_{10}(z_{\rm c}), \theta_{\rm i}\} = \{ 3.89, 2.74\}$ in reference to the best-fit values obtained in \citep{Haridasu:2020pms}, which are similar to the values from the full CMB analysis. 

The second model in the same class\footnote{\href{https: //github.com/flo1984/TriggerCLASS}{https: //github.com/flo1984/TriggerCLASS}} is the New early dark energy (NEDE) \citet{Niedermann:2020qbw, Cruz:2022oqk} model, which provides a similar phenomenology at the background level, however with different perturbations. Indeed, this model has a scalar field with two non-degenerate minimums at zero temperature.  The free parameters governing the phenomenology are the decay time $\Theta_{\rm NEDE} \equiv \{ \log(m/m_0), \fN\}$\footnote{Here $m$ is the mass of the massive scalar field, having $m_0 = 1\, \rm{Mpc}^{-1}$. }, mass and the fraction of early dark energy, respectively. As for the EDE model, here we fix $\log(m/m_0) = 2.58$, following \citep{Niedermann:2020qbw}, varying only $\fN$.

In summary, we assess the physical effects in both the models only through the fraction of early dark energy $(\fEN)$. Apart from the same phenomenological concept (regarding the fact that both models are axion-like scalar field scenarios that introduce additional contribution to the cosmic energy budget in the early Universe), these two models behave differently in alleviating the Hubble tension. Alongside the EDE models we also assess the usual \LCDM model and the non-flat extension with $\Omega_{\rm k} \neq 0 $. 

\subsection{Lensing modelling}
\label{sec:Lensing}
In order to assess the lensing anomaly, the scaling parameter $\Al$, is introduced as a scaling parameter to the lensing potential power spectrum as $C_l^{\psi} \to \Al \times C_l^{\psi}$. This is a phenomenological parameter and when allowed to vary freely in the MCMC analysis and to differ from unity, is interpreted as a lensing anomaly deviation from the \LCDM expectation. Here $\Al$ accounts for both the rescaling of the theory lensing potential power spectrum and the acoustic smoothing of the CMB observables. {In our implementation, however, the $\Al$ parameter accounts only for the amplitude of the acoustic smoothing of the CMB observables but does not rescale the lensing potential power spectrum. This is different from the aforementioned usual implementation in CAMB, where the inclusion of the lensing likelihood to the CMB datasets decreases from $\Al = 1.180 \pm 0.065$ to $\Al = 1.071^{+0.038}_{-0.042}$ \citep{Planck18_parameters}, also having tighter constraints\footnote{Note that these two constraints are $\sim 1.5\sigma$ away, which is a significant deviation with the inclusion of an additional (lensing) likelihood.}. For instance, \citep{Murgia20} have performed an analysis modeling the two effects separately $\{\Al^{\texttt{TTTEEE}},\, \Al^{\phi\phi}\}$, where the former contains the total effect of the acoustic peak smoothing and the latter rescales the theory lensing potential power spectrum. Our implementation corresponds to the ratio of these two parameters and can be easily contrasted by comparing the final constraints $\Al =\Al^{\texttt{TTTEEE}}/\Al^{\phi\phi}$.} {As is well known, this essentially indicates that the rescaling the lensing likelihood itself prefers an even lower level of rescaling, which however in the joint analysis provides tighter constraints converging to a `median' posterior of the two individual likelihoods. } 


On the other hand, following the approach introduced in \citet{Verde17}, one could rescale the lensing potential ($\phi(k,z)$), which yields a similar effect as rescaling the potential power spectrum. However, in our case the rescaling of the potential includes an amplitude ($\Alp$) and the index or tilt ($\nlp$) to a power law as follows:

\begin{equation}
        \phi(k,z) \longrightarrow A_{\rm lp}\left(\frac{k}{k_{\rm lp}}\right)^{n_{\rm lp}}\phi(k,z)\, ,
        \label{eq:lensing}
\end{equation}

here $k_{\rm lp} = 0.1\, h\, \rm Mpc^{-1}$ is the pivot scale which is fixed\footnote{Note that we follow the notation in \citep{Verde17} and we verified that marginalizing on the pivot scale in a range of $k_{\rm lp} \in (0.01 - 0.2)\, h\, \rm Mpc^{-1}$ neither alters the result nor introduces additional correlations.} in the analysis. Note that this formalism was developed to disentangle early cosmology (pre-recombination) from late-time effects, while assuming a general model which describes the early Universe cosmology, as explained in \citep{Vonlanthen10, Audren_2013}, to exclude the effect of the late-time physics on the CMB spectra, it would be enough to assume two new parameters $\{\Alp,\nlp\}$. In order to further reduce the effects of the late-time Universe, we neglect low multipoles of the temperature and polarisation spectra (see figure 5 of \citet{Verde17} ) and set the optical depth value to $\tau=0.01$. Essentially, this implies a marginalization of the lensing potential to rescale accordingly the high-$l$ multipoles of CMB power spectra. Therefore, performing this analysis, we are simultaneously assessing the Early universe constraints for an assumed model (EDE in our case) as we have earlier used in \citep{Haridasu:2020pms} and the constraints on parameters $\{\Alp, \nlp\}$ could aid the discussion on the lensing anomaly. Please refer to \citep{Verde17, Haridasu:2020pms} for a more detailed interpretation of the analysis. {Rescaling the lensing potential in this way is preferable as it also allows us to assess a scale-dependent rescaling, where the effects of the tilt parameter $\nlp$ can mimic the amplitude rescaling effects of $\Alp$. Alongside, the earlier mentioned change in the constraints of the single parameter $\Al$ rescaling both the lensing potential power spectrum and smoothing the CMB spectra is another motivation to study the lensing anomaly while assessing both the effects separately. }

{Note that throughout the analysis presented here, we fix the neutrino sector to the standard scenario i.e., $\sum m_{\nu} = 0.06 \, [\rm{eV}]$ and $N_{\rm eff} = 3.046$. It is also well known that the total mass of the neutrinos will be correlated with the $\Al$, wherein assuming a larger value of $\Al$ will move the limits of the neutrino mass to larger values which we do not investigate in the current analysis. }



\subsection{Data}
\label{sec:Data}
We rely on the standard CMB Planck likelihood\footnote{\href{http://pla.esac.esa.int/pla}{http://pla.esac.esa.int/pla}} \citep{Planck18_parameters} in completeness, which constitutes of the high-$l$ ($l\geq30$) \texttt{TTTEEE} (from \texttt{Plik} likelihood), low-$l$ ($l\leq30$) \texttt{TT} (from \texttt{Commander} likelihood) and low-E \texttt{EE} (from \texttt{SimAll} likelihood) and the Planck-$lensing$, which we hereafter abbreviate as P18. Alongside the P18 dataset, we also use the CMB dataset presented in \citet{Chudaykin20} (hereafter \citetalias{Chudaykin20})\footnote{Note that we do not use the more recent SPT3G \citep{SPT-3G:2021eoc} datasets, which could be an interesting check to perform.}, which was essentially a combination of Planck high-$l$ ($30 \leq l \leq 1000$) \texttt{TT}, SPTPol ($ 50 \leq l \leq 8000 $) \citep{SPT:2017jdf} and SPTLens ($ 100 \leq l \leq 2000$ ) \citep{SPT:2019fqo}. We use the SPTPol likelihood \footnote{\href{https://github.com/ksardase/SPTPol-montepython}{https://github.com/ksardase/SPTPol-montepython}} within \texttt{Montepython} sampler \citep{Brinckmann18,Audren_2013}.  This combination of the data was specifically put forth as one in which the lensing anomaly is not present (i.e, $\Al \neq 1$). Ref \citetalias{Chudaykin20} reports $\Al = 0.990 \pm 0.035$, quoting a very good agreement with the \LCDM expectation. However, we comment later in the \Cref{sec:Results}, further on this inference. For brevity, as for the reference itself, hereafter we abbreviate this dataset combination as C20. In the current analysis, we remain to use only these two data combinations performing no joint analyses with either the late-time LSS datasets or the imposing priors of local estimates of $H_0$. 

As mentioned earlier, for both these models we consider only a one-parameter extension analysis in accordance with the fact that the sampling on the other parameters might provide a clustering of MCMC samples close to $\fEN \to 0$ in the posteriors \citep{Murgia20, Smith:2020rxx, Hill:2020osr, Braglia20}. This is also consistent as we do not include any additional datasets which could alleviate this effect and imply a need to utilize the full parameter space instead \citep{Schoneberg:2021qvd, Ivanov:2020mfr}. Finally, we use the \texttt{getdist}\footnote{\href{https://getdist.readthedocs.io/}{https://getdist.readthedocs.io/}} package \cite{Lewis19} to post-process the chains and infer posteriors.

\section{Results}
\label{sec:Results}

\begin{figure*}
    \centering
    \includegraphics[scale=0.47]{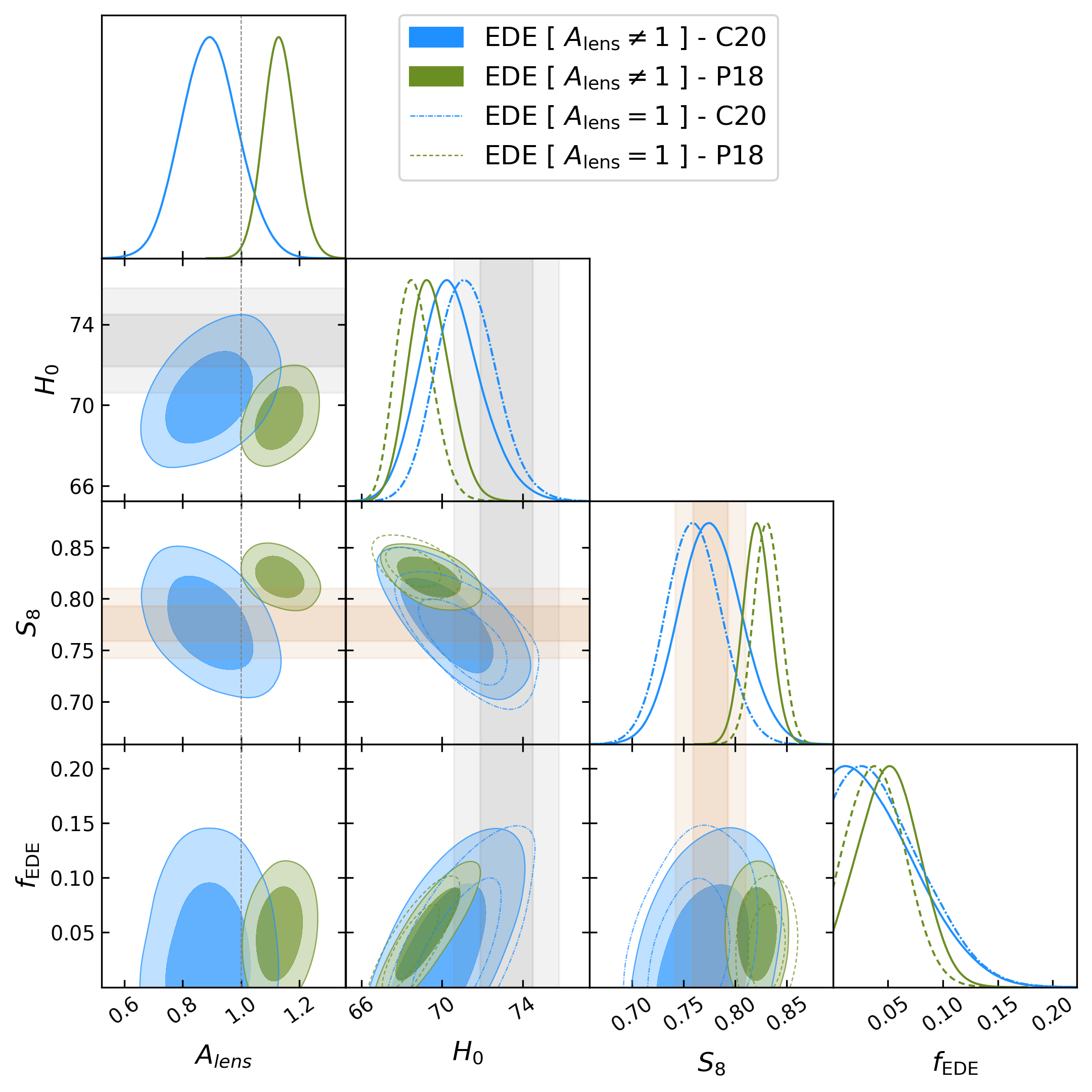}
    \hspace{0.1cm}
    \includegraphics[scale=0.47]{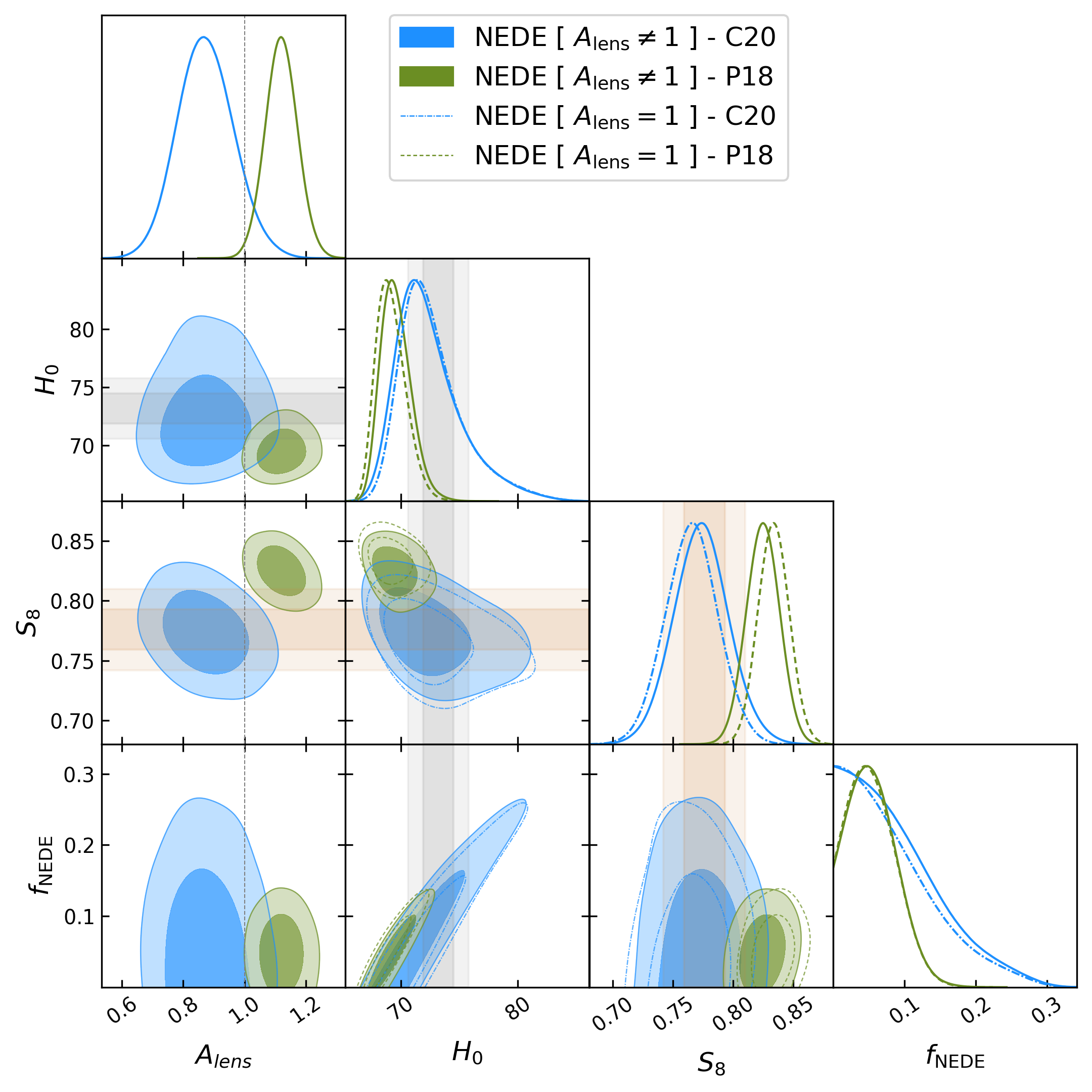}
    \caption{\textit{Left}: Marginalised posteriors for the EDE model with $\Al\neq1$, using the P18 and C20 CMB datasets. The grey band represents the local SH0ES $H_0 = 73.2 \pm 1.3$ \ksM  \citep{Riess:2020fzl, Riess:2021jrx} and the orange band shows the  $S_8 = 0.776\pm 0.017$ \citep{DES:2021wwk}. The grey dashed line marks the $\Al=1$. \textit{Right}: Same as left panel, but for the NEDE model.  }
    \label{fig:EDEandNEDE_Alens_P18vsC20}
    
\end{figure*}

We begin by presenting the results for the correlations between the EDE fractions and the (phenomenological) $\Al$ parameter and then comment on the analysis rescaling the lensing potential. We show the contour plots for the results presented here in \Cref{fig:EDEandNEDE_Alens_P18vsC20} and in \Cref{tab:Constraints_P18}. Once again, we remind that all the results we present for the $\Al$ parameter is accounting only for the additional smoothing and do not rescale the theory lensing potential power spectrum and therefore should not be immediately compared with the constraints in the original \textit{Planck} analysis.

{\renewcommand{\arraystretch}{1.7}%
    \setlength{\tabcolsep}{3pt}%
    \begin{table*}
        \centering
        \caption{We show the constraints for the EDE and the NEDE models using the datasets C20 and P18. We quote $68\%$ C.L. limits for all the parameters, except when we show the $95\%$ C.L. upper limit on $\fEN$. The two columns under each dataset correspond to $\Al$ fixed to unity and $\Al\neq1$, respectively. $H_0$ and $\rd$ are in the units of $\ksM $ and $\rm{Mpc}$, respectively.}
        \label{tab:Constraints_P18}
        \begin{tabular}{c|cccc|cccc}
        \toprule
        Model &\multicolumn{4}{c|}{EDE}&\multicolumn{4}{c}{NEDE}\\
        Data & \multicolumn{2}{c}{C20}& \multicolumn{2}{c|}{P18}& \multicolumn{2}{c}{C20}& \multicolumn{2}{c}{P18}\\ 
            \hline
            &  $\Al=1$ & $\Al \neq 1$&$\Al=1$ & $\Al \neq 1$&$\Al=1$ & $\Al \neq 1$& $\Al=1$ & $\Al \neq 1$ \\
		\hline
		$\omega_{\rm cdm}$            & $0.1181^{+0.0033}_{-0.0047}$ & $0.1193^{+0.0034}_{-0.0049}$ & $0.1232^{+0.0020}_{-0.0025}$ & $0.1232^{+0.0022}_{-0.0026}$ & $0.1207^{+0.0036}_{-0.0065}$ & $0.1217^{+0.0037}_{-0.0066}$ & $0.1245^{+0.0025}_{-0.0034}$ & $0.1238^{+0.0025}_{-0.0034}$\\ 
		$10^{-2}\omega_{\rm b}$& $2.299\pm 0.032            $ & $2.283\pm 0.034            $ & $2.263^{+0.019}_{-0.022}   $ & $2.284\pm 0.024            $ & $2.280\pm 0.027            $ & $2.267\pm 0.028            $ & $2.249\pm 0.016            $ & $2.262\pm 0.018            $\\ 
		$n_s$                       & $0.9855^{+0.0089}_{-0.0089}          $ & $0.9808^{+0.0095}_{-0.0095}          $ & $0.9728^{+0.0059}_{-0.0068}$ & $0.9779^{+0.0065}_{-0.0074}$ & $0.994^{+0.011}_{-0.017}   $ & $0.992^{+0.011}_{-0.017}   $ & $0.9747^{+0.0064}_{-0.0082}$ & $0.9773^{+0.0067}_{-0.0082}$ \\ 
		$H_0$                 & $71.2^{+1.4}_{-1.5}        $ & $70.4^{+1.3}_{-1.6}        $ & $68.63^{+0.8}_{-1.0}       $ & $69.38^{+0.94}_{-1.1}      $ & $72.6^{+1.6}_{-3.2}        $ & $72.4^{+1.7}_{-3.3}        $ & $69.1^{+1.0}_{-1.5}        $ & $69.6^{+1.1}_{-1.5}        $\\  
		$S_8$                         & $0.760\pm 0.027            $ & $0.776\pm 0.030            $ & $0.831\pm 0.013            $ & $0.821\pm 0.013            $ & $0.766\pm 0.022            $ & $0.773\pm 0.023            $ & $0.834\pm 0.013            $ & $0.825\pm 0.014            $ \\
		$r_{\rm d}$              & $145.8^{+2.3}_{-1.4}       $ & $145.7^{+2.3}_{-1.3}       $ & $145.1^{+1.4}_{-1.0}       $ & $144.6^{+1.4}_{-1.2}       $ & $144.4^{+3.6}_{-1.9}       $ & $144.1^{+3.7}_{-2.0}       $ & $144.5^{+1.9}_{-1.3}       $ & $144.5^{+1.9}_{-1.3}       $\\
		$\Al$                         & $---                       $ & $0.894\pm 0.097            $ & $---                       $ & $1.133\pm 0.055            $ & $---                       $ & $0.873\pm 0.092            $ & $---                       $ & $1.121\pm 0.052            $\\ 
		$\fEN$                        & $< 0.119                   $ & $< 0.119                   $ & $< 0.085                   $ & $0.052^{+0.025, \dagger}_{-0.029}   $ & $< 0.211                   $ & $< 0.216                   $ & $< 0.115                   $ & $< 0.115                   $\\ 
		\bottomrule
    \end{tabular}
    \end{table*}
}

\subsection{$\fEN$ vs. $\Al$}

In the left panel of \Cref{fig:EDEandNEDE_Alens_P18vsC20}, we show the marginalised posteriors for the EDE model, using both the P18 (green) and the C20 (blue) datasets, with (solid) and without (dashed) the $\Al$ parameter. Firstly, we notice that the significance of the $\Al$ anomaly remains completely unaltered within the EDE (also NEDE) scenario, with respect to the \LCDM case. This in effect reasserts the conclusion in \cite{Murgia20}, that the lensing anomaly and the existence of a possible early dark energy contribution are independent. However, the correlations between the $\Al$ value and cosmological parameters do affect the limits on the fraction of early dark energy allowed. 

Within the EDE model allowing the $\Al$ parameter to differ from unity allows the constraints from P18 to provide better agreement with both the SH0ES $H_0$ and the LSS $S_8$.  However mild, this indeed is favourable with respect to the fact that alleviating $H_0$ is accompanied by an increase in the value of $S_8$ increasing the tension with the LSS estimate \citep{Ivanov:2020ril, Smith:2020rxx}. While this is a favourable shift in constraints for the P18 data, C20 shows an opposite trend. As we show in \Cref{fig:EDEandNEDE_Alens_P18vsC20}, this effect can be easily understood as a result of correlations between the $\Al$ and $\{H_0,\, S_8\}$ parameters, as larger values of $\Al$ allow for larger values of $H_0$. For the P18 dataset, when allowing $\Al\neq 1$ we find a $95\%$ C.L. limit of $\fE < 0.098$, which is larger than the $\fE< 0.085$, when $\Al$ is fixed to unity. This is an increase of $\sim 15\%$ on the upper limits, implying a larger value for the early dark energy fraction. For the C20 dataset, however, we find only a mild change with the $95\%$ C.L. limit in both cases being almost equivalent at $\fE \lesssim 0.12$. Thus, the limits obtained using the C20 dataset are broad enough that the differences seen with and without accounting for the lensing anomaly are of no major consequences, unlike for the P18 dataset. This is also in accordance with the fact that the lensing anomaly within the C20 dataset is only significant to an order of $\sim 1 \sigma$. We comment more on the lensing anomaly in C20 dataset in \Cref{sec:CommentC20}. The constraint on $H_0$ itself decreases from $H_0 = 71.2^{+1.4}_{-1.5}$ to $H_0 = 70.4^{+1.3}_{-1.6}$, re-increasing the deviation w.r.t to the local estimate \citep{Riess:2021jrx} mildly.

In the right panel of \Cref{fig:EDEandNEDE_Alens_P18vsC20}, we show the same comparison as in the left panel, but for the NEDE model. We note the NEDE model provides slightly larger bounds on the $\fN$ parameter however being almost equivalent to the EDE model for the allowed increase in the $H_0$ itself (see the $5^{\rm th}$ and $9^{\rm th}$ column of \Cref{tab:Constraints_P18}). This shows the ability of P18 dataset to tightly constrain the available physical parameter space in both the EDE and NEDE scenarios. On the other hand, as for the C20 dataset, there exist no tensions, neither for $H_0$ nor for the $S_8$ parameter. While there exists no $S_8$ tension when using the C20 dataset within both the EDE and NEDE cases, the latter is able to alleviate the $H_0$-tension to a much higher degree, making it a more suitable model over the former, at least in this particular comparison. This is in contrast to the recent results presented in \citep{Poulin:2021bjr}, where a combination of Planck and ACTPol data is shown to retain a $\sim 3\sigma$ tension for the NEDE model. As can be seen in \Cref{fig:EDEandNEDE_Alens_P18vsC20}, the NEDE model is also relatively less sensitive to the opening up of the $\Al$ parameter space, essentially displaying a lower degree of correlation with $H_0$. Finally, we summarise the constraints for these analyses in \Cref{tab:Constraints_P18}. We show that the limits on the $\fEN$ parameter do not display large variations, when moving from the  $\Al=1$ to $\Al \neq 1$ scenario, except in the case of the EDE model and the P18 combination. The C20 dataset constraints are large enough to see no major difference in this comparison. The significance of the lensing anomaly in the EDE and NEDE models remains at $\sim 2.4\sigma$ and $\sim 2.3\sigma$, respectively, which is also comparable to the \LCDM case. {Note that in our analysis so far we have always taken into account the lensing dataset, however not rescaling for the theory lensing potential. We infer this as the deviation for the inferred lensing amplitude as the necessary additional acoustic smoothing of the CMB observables when the lensing theory is not altered. Allowing for the rescaling of the lensing potential power spectrum, with and without the inclusion of lensing likelihood would imply $\Al\sim 1.07$ and $\Al\sim1.18$, respectively, as reported in the \textit{Planck} analysis. }
\(\)
\subsection{Early universe analysis}

As we have earlier presented in \citet{Haridasu:2020pms}, here we perform the EUA using both C20 and P18 datasets, having obtained  conclusions for the P18 dataset and the EDE model unaltered from the earlier analysis. For clarity reasons, we report the marginalised posteriors in  \Cref{c1}.
For the C20 dataset and EDE model, we notice a considerably larger reduction in the $95\%$ C.L. upper limits from $\fE < 0.119 $ to $\fE < 0.099$, in comparison to P18 dataset. This is also larger than the variation induced due to the inclusion of $\Al$ as a free parameter and is also accompanied by the reduction in the constrained value of $H_0 = 71.2^{+1.4}_{-1.5}$ to $H_0 = 69.5\pm 1.6$. While this is an anticipated result in accordance with our previous work, interestingly we also find that the constraints on the spectral index $n_{\rm s}$ are now in better agreement with those obtained using the P18 dataset. The EUA analysis brings down the mean values of the $n_s$, however having a larger elongated tail for larger values. Note that the shift in the values of $n_s$ is now brought about only due to the marginalization over the lensing potential (see \Cref{eq:lensing}) and is not a consequence of excluding the Planck \texttt{TT} dataset. 


In \Cref{fig:EDEandNEDE_EUA_P18vsC20} we show the contour plots for both the EDE and the NEDE models. Notice that the posteriors of $n_s$ in the case of EUA for the EDE model are now more in agreement with the values from P18 and \LCDM model, shifting from $n_s = 0.986\pm 0.009$ to $n_s = 0.974\pm 0.011$ yet having  similar uncertainties. This shift is at a level of the standard deviation for each of the constraints. This in turn indicates that the discrepancies between the P18 and the C20 datasets can be to some extent reconciled within the EDE scenarios in the EUA, which is once again in favor of the model. As for the NEDE model, we see a very similar behavior still retaining a larger posterior parameter space, and mildly better than the EDE model when using the P18 dataset. For the C20 dataset, however, as expected the limits on $\fN$ are lowered yet remain large enough to be able to alleviate the $H_0$-tension and is of no major consequence for the intended purpose of the model. Therefore once again performing slightly better than the EDE model. {In addition, allowing for the $\Al$ freedom within the EUA analysis also provides a better agreement between the constraints obtained from the P18 and C20 datasets, as shown in \Cref{fig:Lensing_P18vsC20}.}

More recently \citep{Mortsell:2021nzg} revised the local $H_0$ value taking into account the systematics within the color-intensity relation derived using the Cepheid color measurements and brought down the values of $H_0$ to complete agreement or reducing the current $\sim 5\sigma$ tension to utmost $2.7\sigma$. In essence, taking this estimate at face value implies either no modification or only a mild shift from \LCDM. However, the latter case of mild tension now implies that a large number of modifications can perform similarly and that a small fraction of early dark energy can be sufficient.

\subsection{Comments on lensing anomaly}
In this section, we report on the CMB lensing analysis that we perform with the additional degrees of freedom attributed to the marginalization of the lensing potential with and without the inclusion of the $\Al$ parameter.

As described in \Cref{sec:Lensing}, when performing the EUA we marginalize on the rescaling parameters $\{\Alp\,,\nlp\}$ of the Newtonian lensing potential. We now perform the EUA with the $\Al$ parameter free for the case of \LCDM model and then extend to EDE/NEDE models as well. Firstly, we find that the individual marginalised posteriors of $\Alp = 1.035^{+0.038}_{-0.043}$ and $\nlp =-0.020\pm 0.025$, are very well consistent with the expectation. However, we find that in 2 dimensions, within this comparison there exists an apparent $\sim 3.6 \sigma$\footnote{We compute this deviation by constructing a multivariate probability distribution function for $\{\Alp\,, \nlp \}$ using the individual posteriors and the corresponding correlation. Then estimate the distance from $\{1, 0\}$. The correlation between the $\Alp$ and $\nlp$ is $\sim 0.89$. We thank  Marco Raveri, for having pointed this out.} deviation from the usually fixed values of $\{\Alp\,,\nlp\} = \{1, 0\}$ for the P18 dataset. {This apparent deviation is clearly evident due to the fact that we fix the $\tau = 0.01$ in the EUA and will be relaxed if $\tau$ is allowed to vary freely, however providing only an upper limit. We verify this by including low-$l$/lowE one at a time to EUA, while allowing $\tau$ to be a free parameter as shown in \Cref{fig:Lensing_P18EUA}. }

\begin{figure}
    \centering
    \includegraphics[scale=0.47]{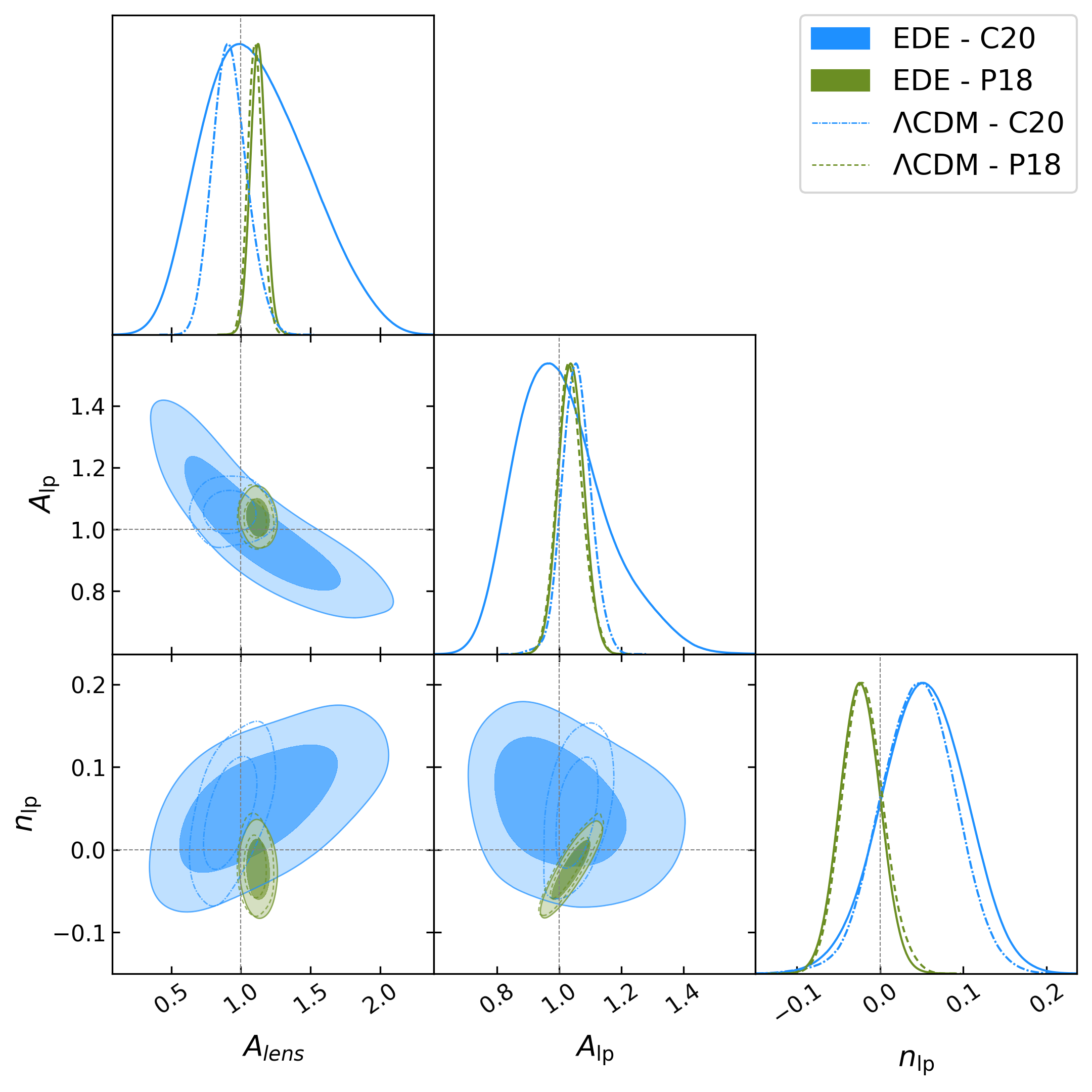}
    \caption{We show the correlations between various lensing rescaling parameters for the EDE and the \LCDM models, using both the P18 and C20 datasets.  The dashed lines mark the scenario with no deviation from the standard expectations, $\{\Al,\Alp, \nlp\} = \{1,1,0\}$. }
    \label{fig:Lensing_P18vsC20}
\end{figure}

\begin{figure}
    \centering
    \includegraphics[scale=0.47]{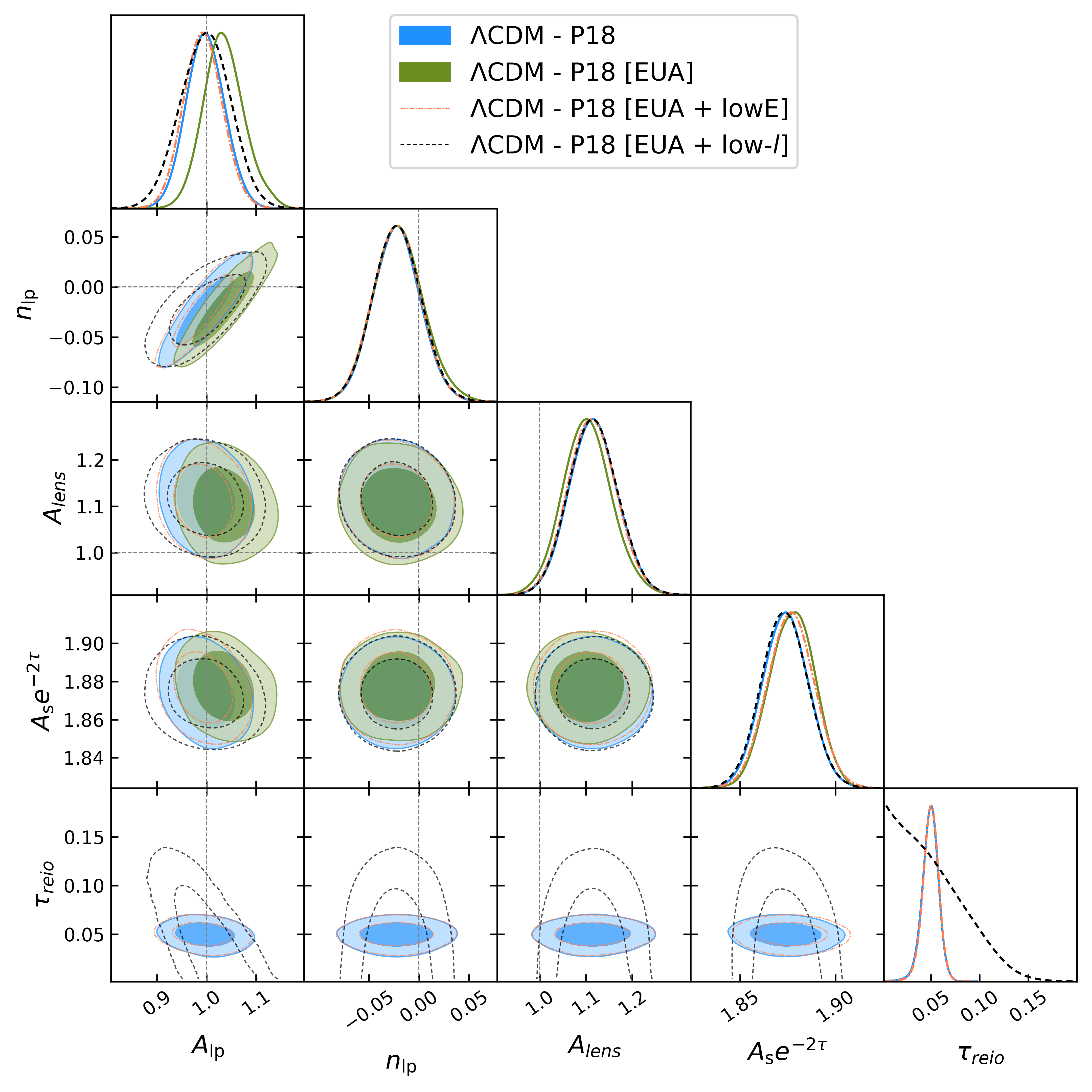}
    \caption{We show the correlations between lensing rescaling parameters for the \LCDM models, using both the P18 EUA and combinations of low-$l$ and /or lowE data.  The dashed lines mark the scenario with no deviation from the standard expectations, $\{\Al,\Alp, \nlp\} = \{1,1,0\}$. }

    \label{fig:Lensing_P18EUA}
\end{figure}

Note that this shift in  the parameter space not only implies a rescaling of the amplitude but also a scale-dependence, while not being a replacement or recasting the $\Al$ anomaly in terms of $\{\Alp\,,\nlp\}$, as the constraints on $\Al$ remain the same as in the complete CMB dataset analysis. This is expected, as the $\Al$ parameter in our analysis is only accounting for the additional acoustic smoothing but does not rescale the theory lensing potential. In \Cref{fig:Lensing_P18vsC20}, we show the contours for the lensing potential rescaling parameters and the $\Al$ for the EDE and \LCDM models.  

Assessing the effects of lensing anomalies within the EDE models, earlier \citep{Murgia20} have performed an analysis with two distinct parameters to marginalize the CMB lensing in P18 dataset. These two parameters $\{\Al, \Al^{\phi\phi}\}$ are modeled to rescale the smoothing of acoustic peaks and to rescale the lensing potential power spectrum, respectively. These two parameters are equivalent to the $\{\Al, A_{\rm lp}^2\}$\footnote{Note that in the usual practice allowing $\Al\neq1$, rescales the lensing potential as $A_{\rm lp} \sim \sqrt{\Al^{\phi\phi}}$, when the tilt $\nlp=0$ is fixed. } in our notation, however, here in addition we also have the tilt parameter $n_{\rm lp}$, which is fixed to zero in \citep{Murgia20}. We verify with the P18 dataset that $\nlp = 0$ in our analysis will reproduce the results in \citep{Murgia20} (contrast TABLE VI therein and our \Cref{fig:Lensing_P18vsC20}). 

{Clearly the aforementioned apparent $\gtrsim 3 \sigma$ deviation in the EUA analysis is equivalently found also for the EDE and NEDE models, see \Cref{fig:Lensing_P18vsC20}. However, when the low-$l$ \texttt{TT} and low-E datasets are included, while allowing the lensing potential to be rescaled, this deviation reduces to the level of $\sim 1\sigma$. Once again, we assert that this deviation is completely driven by the fact that we set the reionization depth $\tr = 0.01$. However, while doing so, the cosmological parameters and the inferences for the additional smoothing of the CMB spectra do not change. This simultaneously serves the purpose of validating the EUA, which is unaffected by the assumed values of $\tr$, and illustrating the correlations between the reionization physics and the theory lensing potential expectation. In \Cref{fig:Lensing_P18EUA}, we show the corresponding contours.  } 

{To assess this issue further, we perform the analysis rescaling the lensing potential and the lensing power spectrum in several combinations of CMB datasets with and w/o the inclusion of either low-$l$ and lowE datasets. In essence, we find that the tensions between the CMB observable and the LSS lensing observables \citep{Smith:2020rxx, Joudaki17a, DES:2021wwk} can all be attributed to the tight constraints on the value of the $\tr$ placed by the lowE dataset. Note that this has no role to play as an argument in the discussion regarding the $H_0$-tension. For instance, a mildly lower value of $\tr$ can aid to decrease the values of $A_{\rm s}$ (consequently $\sigma_8$), with important implications for  the $\sim 2\sigma$ deviation in the values of $A_{\rm s}$ estimated from the LSS and CMB observables \citep{Smith:2020rxx}. In this context, on the one hand, models that are able to modify the reionization history would provide interesting avenues to test for either improving the agreement between the LSS and CMB datasets. And on the other hand, this difference when enhanced within a given model can be used for model selection, needless to say, within the \LCDM model we already have a $\sim 2 \sigma$ discrepancy. } 

{The C20 dataset though is completely consistent with the expectation of $\{A_{\rm lp}\,,n_{\rm lp}, \Al\} = \{1, 0, 1\}$ at $\lesssim 1 \sigma$, however showing mild correlations between the  parameters. The significance of the deviation of $\Al$ form unity is already reduced within the \LCDM model and is completely unseen when extending to the EDE model, also showing the degeneracy between the $\Al, \Alp$ parameters. In \Cref{fig:Lensing_P18vsC20}, one can clearly notice the correlation between the parameters $\nlp$ and $\Al$ for the \LCDM model, indicating that a scale-dependent tilt of the lensing potential can mimic the effects of amplitude scaling $\Al$.  }


\section{Summary}
\label{sec:Conclusions}
In this work, we have assessed a few subtleties in relation to the EDE/NEDE models and the lensing anomaly contrasting against two different CMB datasets. Our main conclusions can be summarized as follows:
\begin{itemize}
    \item While the lensing anomaly (yet mild) is not correlated with the fraction of EDE/NEDE it  does correlate with the cosmological parameters and yield subtle differences in the inferred limits of $\fEN$. Also, we find  that contrasting the EDE/NEDE against the early universe information with CMB datasets, indeed reduces the limits for the allowed $\fEN$ values.
    \item NEDE model provides a larger posterior parameter space in comparison to the EDE model, being more suitable in alleviating the $H_0$ and $S_8$ tensions. 
    \item Rescaling the theory lensing potential (see \Cref{eq:lensing}) and accounting for the additional smoothing in the CMB spectra we assert that an anomaly for $\Al \sim 1.1$ at a $\sim 2 \sigma$ confidence level should be addressed by modifications to the pre-recombination physics, within the P18 dataset. 
    \item {However, using the CMB dataset (C20) considered, we find that the mild additional acoustic smoothing of the CMB spectra can be addressed by a scale-dependent rescaling of the lensing potential. }
\end{itemize}

Our results also corroborate the findings of \cite{Fondi:2022tfp}, where no additional statistical evidence in favor of EDE models was found when allowing $\Al \neq 0$ and $\Omega_{\rm k} \neq 0 $ simultaneously, indicating that the mild deviations of $\Omega_{\rm k}<0$ and $\Al>1$ in the \textit{Planck} CMB data do not result in a higher preference for EDE modifications. {While we have not studied the same in EDE/NEDE models, our analysis here shows within the \LCDM model with $\Omega_{\rm k} \neq 0$, that the additional smoothing of the CMB spectra remains with $\Al>1$, while the amplitude of the lensing potential is degenerate with the curvature of the universe. } Finally, we also comment on the amount of lensing anomaly in the C20 dataset explicitly. 



\section*{Acknowledgements}
BSH, HK and MV are supported by the INFN INDARK grant. HK acknowledges support for this work from NSF-2219212. The authors are thankful to Marco Raveri, Guillermo F. Abellan, and Carlo Baccigalupi, Riccardo Murgia for useful discussions at different stages of the work. We are also grateful to Anton Chudaykin for early correspondence and elaborations on their work assessing the lensing anomaly in the C20 dataset.

\section*{Data availability}
The data utilized in this article are all publicly available, and appropriately referred to within the text (see  \Cref{sec:Data}).

\appendix




\bibliographystyle{mnras}
\bibliography{bibliografia} 



\appendix

\section{Lensing anomaly in the C20 dataset}
\label{sec:CommentC20}
\begin{figure}
    \centering
    \includegraphics[scale=0.47]{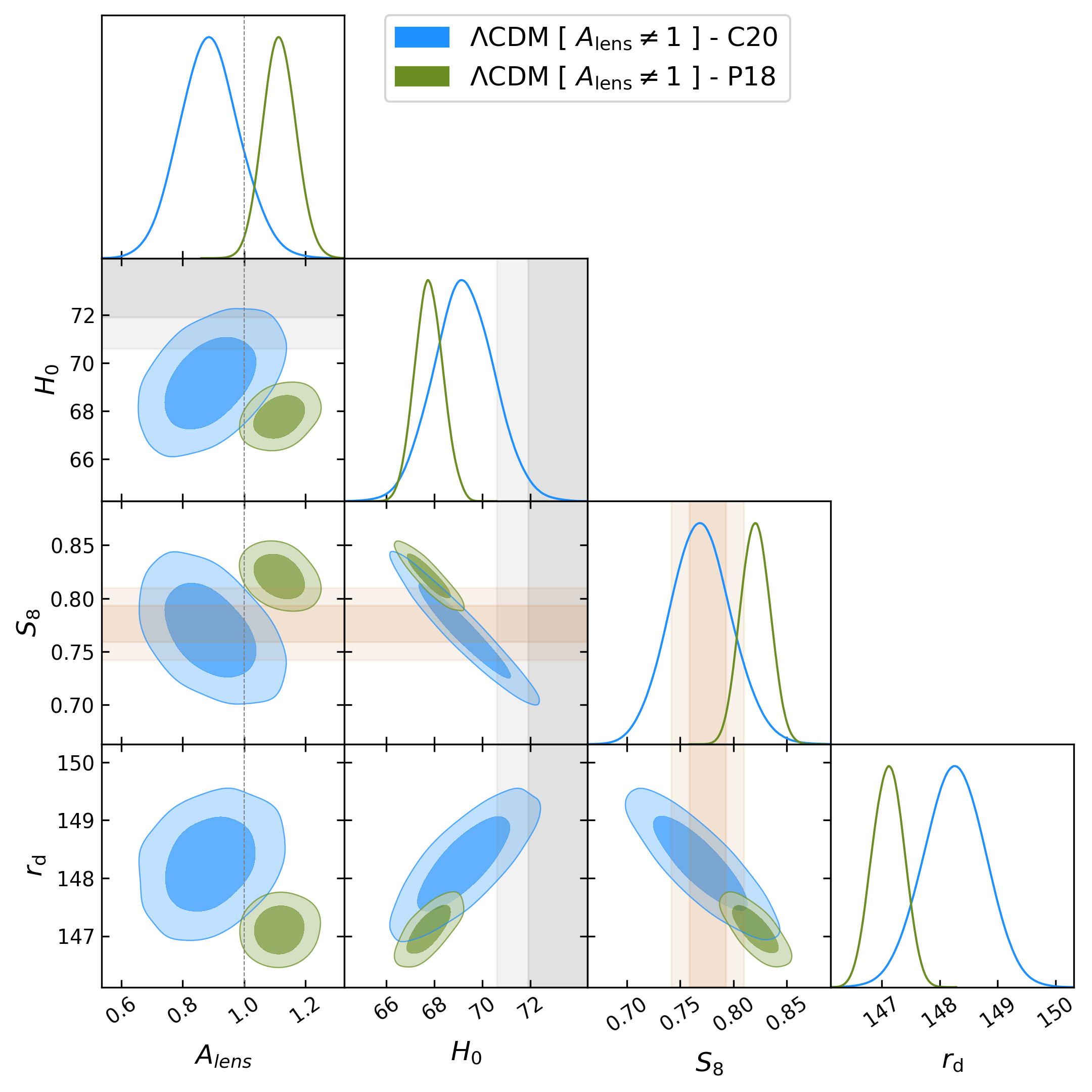}
    \caption{Marginalised posteriors for the \LCDM model with $\Al\neq1$, using the P18 ( high-$l$ [\texttt{TTTEEE}] + low-$l$ [\texttt{TT}] + low-E [\texttt{EE}] + Planck-$lensing$ ) and C20 ( high-$l$ [\texttt{TT} ($l<1000$)] + low-$l$ [\texttt{TT}] + low-E [\texttt{EE}] + SPTPol + SPTLens ) CMB datasets. The grey band represents the local SH0ES $H_0 = 73.2 \pm 1.3$ \ksM  \citep{Riess:2020fzl} and the orange band shows the  $S_8 = 0.776\pm 0.017$ \citep{DES:2021wwk}. The grey dashed line marks the $\Al=1$. }
    \label{fig:Alens_P18vsC20}
\end{figure}
We reanalyze the \citetalias{Chudaykin20} dataset, firstly noticing that the parameter chosen to assess the lensing anomaly in this analysis is disparate from the traditional parameter. The analysis in \citetalias{Chudaykin20} utilizes the amplitude of the lensing potential ($A_{\rm lp}$) instead of the phenomenological $\Al$ which alters lensing potential power spectrum as described in \Cref{sec:Theory}. We indeed verify that when $\Al$ is fixed to unity, we obtain $A_{\rm lp} = 0.990 \pm 0.035$, replicating the analysis in \citetalias{Chudaykin20}. To assess the actual value of the lensing anomaly, we now perform the analysis appropriately fixing $A_{\rm lp} =1$ and find $\Al = 0.88\pm 0.10$, which is very much in agreement with the value $\Al = 0.81\pm 0.14 $ quoted in \citep{SPT:2017jdf}. While rest of the parameters are in very good agreement with those quoted in Table 1. of Ref.~\citetalias{Chudaykin20}. We, therefore, infer that the C20 dataset combination is \textit{not} completely free of the lensing anomaly but reproduces the deviation as originally seen in the SPTPol data. This unfortunately deters the primary motivation presented in \citetalias{Chudaykin20} to construct a dataset free of the lensing anomaly. Nevertheless, the significance of deviation in the C20 dataset is milder only at $\sim 1.2 \sigma$. Also, the value of $\Al$ we find using the C20 dataset is $~2.1 \sigma$ away from the P18 estimate, which is of mildly less significance than the $\sim 2.9\sigma$ between SPTPol-only and Planck \texttt{TT} reported in \citep{SPT:2017jdf}. We, therefore, utilize the C20 dataset as an alternate CMB data combination, to assess the further variations on the limits for $\fEN$, while taking into account the lensing anomaly and those that could be seen in an EUA. We presented the results obtained in the EUA and later proceed with the comments on the role of lensing anomaly in assessing the $\fEN$ limits.

In \Cref{fig:Alens_P18vsC20}, we show a comparison of the constraints obtained from the P18 and C20 datasets. Clearly, these two CMB datasets do not completely agree\footnote{As the P18 and C20 have low-$l$ \texttt{TT}  and low-E \texttt{EE} and high-$l$ \texttt{TT} ($l< 1000$) in common, we do not immediately report the significance for tensions, however, it is clear that the two datasets show different behavior when the $\Al\neq 1$ freedom is allowed.}, having a discrepancy of at least $\sim 2\sigma$ and more, when $\Al\neq 1$ is allowed. This is in contrast to the $\leq 2\sigma$ deviation reported in \citetalias{Chudaykin20}, when $\Al = 1$, while we agree to the assessment that the $S_8$ is lower and $H_0$ is higher, already alleviating the tensions to some extent. As can be seen in \Cref{fig:Alens_P18vsC20}, allowing $\Al$ to be different from  unity does not immediately imply larger values of $H_0$, but extends the posteriors towards lower values for the C20 and higher values for P18 datasets, respectively. This effect is due to the positive correlation of the order $\sim 0.44$ ($\sim 0.30$) between the $\Al$ and $H_0$, for the C20 (P18) dataset. A similar level of anti-correlation is observed between the $\Al$ and $S_8$ parameters. And in turn, this clearly indicates the need to assess the same in extensions to \LCDM, intended to address the $H_0$-tension {as we have elaborated in the main text}. 


\section{Lensing anomaly and Curvature ($\Omega_{\rm k}\neq 0$)}

\label{sec:closed_universe}

\begin{figure}
    \centering
    \includegraphics[scale=0.47]{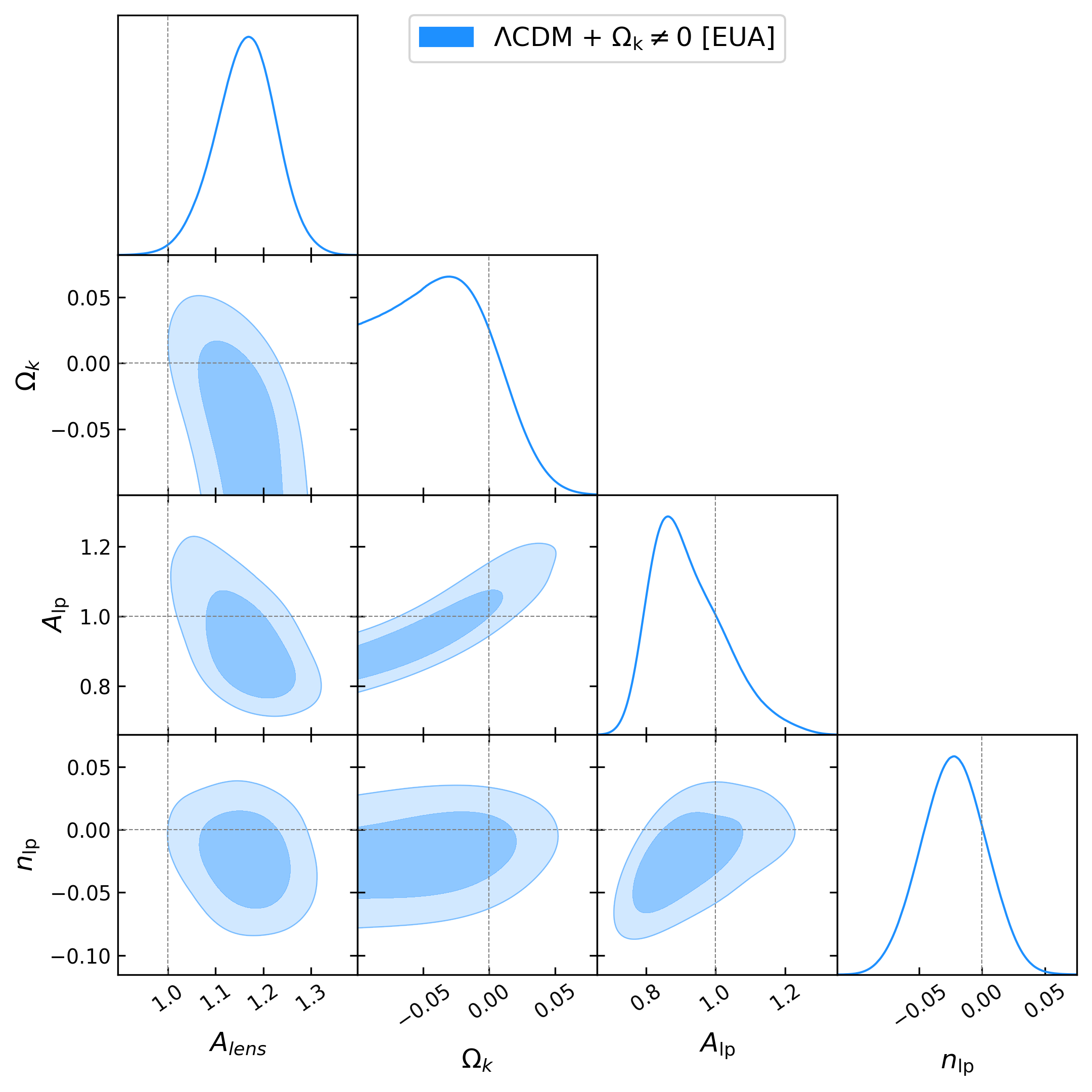}
    \caption{Marginalised posteriors for the k\LCDM model with $\Al\neq1$ and rescaling the theory lensing potential, using the \textit{Planck} high-$l$ [\texttt{TTTEEE}] + $lensing$. Note here, in our analysis $\Al$ is only an additional smoothing on the CMB spectra and does not rescale the lensing potential}
    \label{fig:Alens_OmegakP18}
\end{figure}

{It has been shown that some correlation between the lensing anomaly and the curvature exists and that this could imply a crisis if the anomaly were to indicate and hence be resolved within a  closed universe $\Omega_{\rm k} < 0$ \citep{DiValentino:2019qzk} (see also \citep{Hazra:2022rdl}). In line with our EUA we repeat the same analysis leaving both the $\Al$ (accounting only for the additional smoothing, as is our implementation) and $\Omega_{\rm k}$ as free parameters. Here EUA already implies that the theory lensing potential is being rescaled. We show the marginalized contours for this analysis in \Cref{fig:Alens_OmegakP18}, and find no strong correlation between our $\Al$ parameter and the curvature. As can be clearly seen, the curvature can clearly compensate for the lensing amplitude which however rescales the theory lensing potential, but the additional smoothing given by $\Al$ in our implementation is independent and presents no correlation. This once again reasserts that two distinct contributions exist to the total anomalous smoothing of the CMB spectra and that two different corrections might be necessary to fully explain away the lensing anomaly in the standard scenario. }

{In comparison to the \LCDM case, we find that the constraints on the $\Al$ parameter are driven mildly towards larger values accommodating for the low values of $\Alp$, which in turn is due to the strong negative correlations of the latter with $\Omega_{\rm k}$. However, in this case, the $\Al$ is only going to be larger than unity providing no compensation for the $\Omega_{\rm k} < 0$. } {This result is also in agreement with the recent assessment in \cite{deCruzPerez:2022hfr}, where it has been shown that the $\Al>1$ extension is strongly preferred over spatially curved hypersurfaces ($\Omega_{\rm k}$ < 0 ). }

\section{Triangle plots of EDE/NEDE in the Early Universe Analysis}
\label{sec:Tri_figures}
\label{c1}
We show the marginalized posteriors in the early universe analysis of the EDE and NEDE models utilizing both the P18 and C20 datasets. 


\begin{figure*}
    \centering
    \includegraphics[scale=0.47]{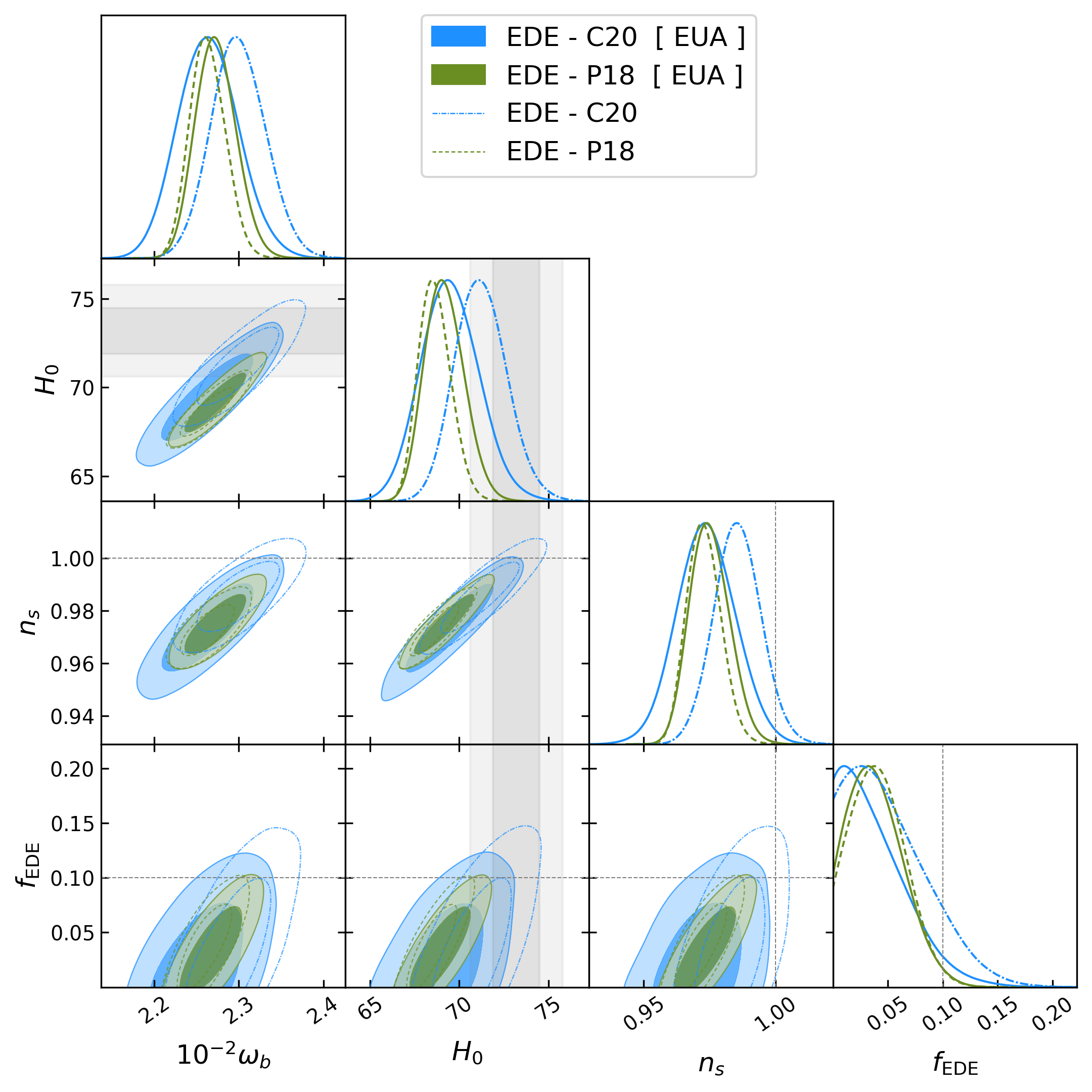}
    \hspace{0.1cm}
    \includegraphics[scale=0.47]{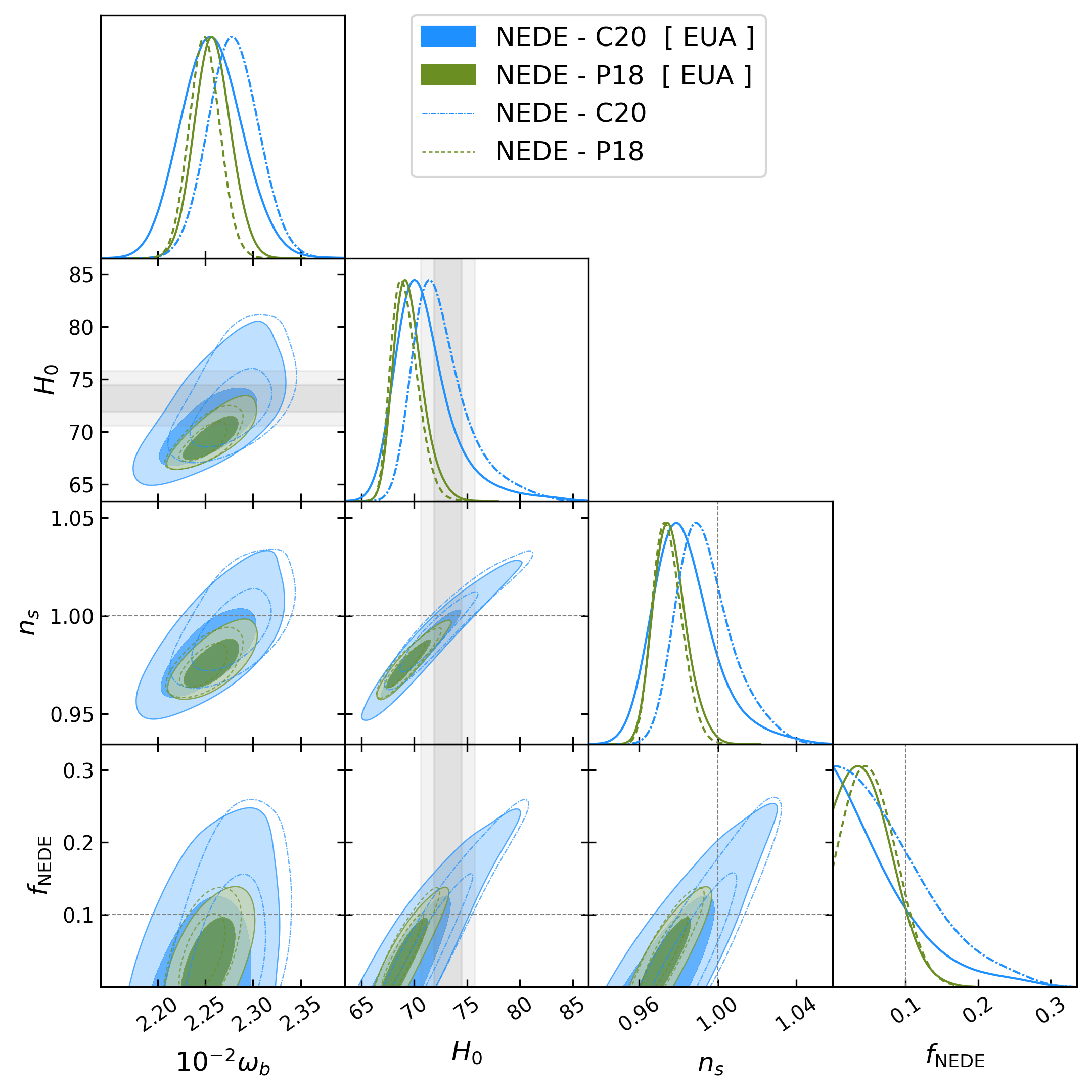}
    \caption{\textit{Left}: Marginalised posteriors for the EDE model with using the P18 and C20 CMB datasets in the EUA. The grey band represents the local SH0ES $H_0 = 73.2 \pm 1.3$ \ksM  \citep{Riess:2020fzl}. We mark with the grey dashed lines $\fE = 0.1$ and $n_s = 1.0$. \textit{Right}: Same as the left panel, but for the NEDE model.  }
    \label{fig:EDEandNEDE_EUA_P18vsC20}
\end{figure*}

\bsp	
\label{lastpage}
\end{document}